\documentclass[12pt]{article}
\usepackage{amsmath}
\usepackage{amssymb}
\usepackage{amsfonts}
\usepackage{epsfig}

\begin{document}

\title{Can gravitational collapse sustain singularity-free trapped surfaces?}
\author{Manasse R. Mbonye$^{1,2}$ and Demos Kazanas$^{2}$ \\
%EndAName
\ \ \ \ \\
$^{1}$\textit{Department of Physics, }\\
\textit{Rochester Institute of Technology, }\\
\textit{84 Lomb Drive, Rochester, NY 14623.} \ \\
\ \ \ \ \ \ \ \ \ \ \\
$^{2}$\textit{NASA/Goddard Space Flight Center,}\\
\textit{Mail Code 663, Greenbelt, MD 20785}}
\date{ }
\maketitle

\begin{abstract}
In singularity generating spacetimes both the out-going and in-going
expansions of null geodesic congruences $\theta ^{+}$ and $\theta ^{-}$
should become increasingly negative without bound, inside the horizon. This
behavior leads to geodetic incompleteness which in turn predicts the
existence of a singularity. In this work we inquire on whether, in
gravitational collapse, spacetime can sustain singularity-free trapped
surfaces, in the sense that such a spacetime remains geodetically complete.
As a test case,\ we consider a well known solution of the Einstien Field
Equations which is Schwarzschild-like at large distances and consists of a
fluid with a $p=-\rho $ equation of state near $r=0$. By following both the
expansion parameters $\theta ^{+}$ and $\theta ^{-}$ across the horizon and
into the black hole we find that both $\theta ^{+}$ and $\theta ^{+}\theta
^{-}$ have turning points inside the trapped region. Further, we find that
deep inside the black hole there is a region $0\leq r<r_{0}$ (that includes
the black hole center) which is not trapped. Thus the trapped region is
bounded both from outside and inside. The spacetime is geodetically
complete, a result which violates a condition for singularity formation. It
is inferred that in general if gravitational collapse were to proceed with a 
$p=-\rho $ fluid formation, the resulting black hole may be singularity-free.
\end{abstract}

\section{Introduction}

The concept of a black hole as a gravitationally collapsed object has
effectively been with us since 1916 when Schwarzschild first presented a
solution to the Einstein field equations for the gravitational field of a
spherically symmetric mass. To date, the physics of these end-products has
evolved both in breadth and depth, both theoretically and observationally,
so much so that black holes are currently considered virtually discovered.
While the horizon of a black hole (a 2-sphere coordinate singularity)
provides the classical observational limit with regard to the dynamical
evolution of the hole, it is widely believed that during spacetime collapse
to form a black hole, the matter itself continues to implode unimpeded
beyond the horizon towards a physical singularity. Such perception has been
motivated, from a theoretical view point, by the apparent absence of any
known force that would otherwise counteract the action of gravity under the
circumstances.

Lately, the issue of whether or not spacetime singularities do exist is
being revisited [1]. There are two main reasons for this. First there is the
need to deal with the information loss paradox in black holes [2]. Further,
modern frameworks (like string theory) attempting to formulate a quantum
theory of gravity assume the existence of fundamental length scale and thus
seek singularity free spacetimes. There have been suggestions and
speculations that singularity free spacetimes may actually be generic in
nature [3]. Indeed the concept of non-singular spacetimes actually dates
back quite a while. For example, Einstein [4] invoked a cosmological
constant in his field equations in order keep the universe in equilibrium
from collapse against its own matter-generated gravity. Later, [5] Sakharov
considered the equation of state for a superdense fluid of the form $p=-\rho 
$ and Gliner suggested [6] that such a fluid could constitute the final
state of gravitational collapse. Currently, researchers\ continue\ to
investigate the possibility of non-singular gravitational collapse.. For
example, Dymnikova (see [7] and refs. therein) has constructed a solution of
the Einstein field equations for a non-singular black hole containing at the
core a fluid $\Lambda _{\mu \nu }\left( r\right) $ with anisotropic pressure 
$T_{0}^{0}$ $=T_{1}^{1}$, $T_{2}^{2}=T_{3}^{3}$. Another line of
investigation in this area initiated by Markov [8], suggests a limiting
curvature approach. This idea has been explored further\ by several authors,
see for example [9] [10] [11] [12] [13] [14]. The notion of non-singular
collapse does lead (among other things) to some interesting speculations
such as universes generated from interior of black holes [15] [16] [17].
Non-singular collapse has also been extended to considerations of other
modified forms of gravitational collapse such as boson stars [18] and
gravitational vacuum stars (gravastars) [19].

A common feature in all these treatments is that the geometry of the
spacetime in question is Schwarschild at large $r$ and de Sitter-like at
small $r$\ values. Issues to do with direct matching of an external
Schwarzschild vacuum to an interior de Sitter one have previously been
discussed [20] from the point of view of the junction conditions [21]. In
fact this approach has been employed to suggest modifications in some
treatments [22]. If gravitational collapse is to result into a non-singular
object with a de Sitter-like core then junction constraints imply the matter
field profile between or across the Schwarzschild/de Sitter boundary should
have a radial dependent equation of state of the form $1\lesssim
w(r)\lesssim -1$\ that smoothly changes the matter-energy from a stiff fluid
\ to a de Sitter-like fluid. In particular, inside the black hole where the
radial coordinate becomes timelike such a field could take on a
quintessential time dependant character, probably evolving as $0\lesssim
w(r)\lesssim -1$.

According to general relativity the 2-spheres inside of and concentric to
the event horizon form a family of trapped surfaces $\Sigma $, in that the
null geodesics of both ingoing and outgoing families, orthogonal to each
such a 2-surface, are converging. This means that the tangent vector field $%
l^{\alpha }$ of any hypersurface-orthogonal null geodesic defined on any
such 2-surfaces have a negative expansion,\ $\theta =$ $l_{~;\alpha
}^{\alpha }$ $<0$. Thus spacetimes with horizons should contain trapped
surfaces. In the 1960s, inquiries in the nature of spacetime singularities
by Penrose and Hawking led to a quantitative theoretical description of
gravitational collapse through the enunciation and proof of the famous
Singularity Theorems [23] [24]. These studies were also extended by Hawking
and Ellis [25] towards the concept of an initial singularity in cosmology.
What the singularity theorems proved was that, subject to certain energy
conditions, (see also Section 2.2) a closed trapped surface will contain a
singularity. The question then is whether the geometry resulting from
gravitational collapse can sustain trapped surfaces without a singularity.
This is a reasonable question especially in view of the recent findings by
Ellis about a related phenomenon in cosmology [26], namely that cosmic
dynamics can allow closed trapped surfaces without leading to an initial
singularity. In this paper we investigate whether it is in any way possible
to have a family of trapped surfaces $\Sigma _{T}$, resulting, say, from
spherically symmetric gravitational collapse, that contain no
future-directed singularity.

The rest of the paper is organized as follows. In section 2 we briefly
review the ideas of convergence of null geodesics and how they lead to the
concept of trapped surfaces. Section 3 introduces the general class of
spacetimes under consideration and discusses suitable coordinates. In
section 4 we compute the expansion of both the out-going and in-going null
geodesics in a well known non-singular spacetime that is an exact solution
of Einstein Equations. We use the results to demonstrate an unusual
character that inside such a black hole there is a region \ $0<r<r_{0}$ for
which $\theta ^{+}\theta >0$. The result violates conditions for the
existence of a physical singularity. In section 5 we provide a summary and
conclude the paper.

\section{ Null geodesics and gravitational collapse}

In this section we briefly review two concepts to be used in the forthcoming
discussion, namely the concept of geodesic convergence and the concept of
trapped surfaces.

\subsection{Convergence}

The propagation of timelike or null rays in a given geometry is governed by
the Raychaudhuri equation, 
\begin{equation}
\frac{d\theta }{dv}=\varkappa \theta -\left( \gamma _{c}^{c}\right)
^{-1}\theta ^{2}-\sigma _{\alpha \beta }\sigma ^{\alpha \beta }+\omega
_{\alpha \beta }\omega ^{\alpha \beta }-R_{\alpha \beta }l^{\alpha }l^{\beta
},
\end{equation}%
where $\theta $ is the expansion rate, $\sigma $ is the shear rate and $%
\omega $ is the twist. $R_{\alpha \beta }$ is the Ricii tensor. Further, $%
\gamma _{c}^{c}$ is the trace of the projection tensor for null geodesics
while $\varkappa $ is identified with the surface gravity. For simplicity of
treatment, we shall consider a twistless $\omega =0$, shearless, $\sigma =0$
spacetime that is type D in the Petrov classification [27]. Then the
geodesic tangent vector can be chosen to be in the principal null direction.
Use of the Einstein Equations implies 
\begin{equation}
\frac{d\theta }{dv}+\frac{1}{2}\theta ^{2}+\kappa \left( T_{\alpha \beta }-%
\frac{1}{2}g_{\alpha \beta }T\right) l^{\alpha }l^{\beta }=0.
\end{equation}%
where $\kappa =8\pi G$. Eq. 2 shows that the expansion/convergence $\theta $
of a geodesic congruence will, among other things, depend on energy
conditions. In particular, for convergence of \ a null geodesic congruence,
it is necessary (and sufficient) that fields influencing the geometry
satisfy the weak energy condition, i.e. $R_{\alpha \beta }k^{\alpha
}k^{\beta }\geq 0$ which leads to $\rho +p\geq 0$. The expansion $\theta $
of the congruence (or the evolution of the principal null vector $l^{\alpha
} $) in a given geometry $g_{\alpha \beta }$ is defined by 
\begin{equation}
\theta =l_{;\alpha }^{\alpha }=\frac{1}{\sqrt{-g}}\partial _{\alpha }\left( 
\sqrt{-g}l^{\alpha }\right) .
\end{equation}%
Thus with the knowledge of the metric, one can construct $\theta $.

\subsection{Trapped surfaces and singularities}

Consider a spacelike 3-hyperface $\Sigma $ on which one has defined a metric 
$\gamma _{ij}$ and an extrinsic curvature tensor $K_{ij}$. The space $\Sigma 
$ can be foliated with a family of 2-surfaces, $S$ on which one can define
hyperface-othorgonal null vectors $l^{\alpha }$. There are two such families
of vectors defined on a surface $S$, namely in-going and out-going rays. The
expansion of the in-going congruence is always converging, while that of the
out-going one can either diverge or converge depending on the circumstances.
In the case that the expansion of both families of these tangent vectors are
negative, $\theta <0$, then the surface $S$ is said to be a trapped surface $%
S_{T}$. There will be a marginally trapped surface(s) $S_{Ma}$ in $\Sigma $
for which the expansion of the outgoing null geodesics defined on each point
on $S_{Ma}$ are vanishing, $\theta =0$. For black holes in the process of
formation, this surface is the apparent horizon. It defines the outer
boundary of the family of closed trapped surfaces $S_{T}$ in this geometry.
The existence of trapped surfaces in black holes (along with other
conditions) was used by Penrose and Hawking to predict that such spacetimes
should have singularities. Briefly, the main argument goes something like
this [26] [28]. A given spacetime will contain at least one incomplete
geodesic, provided:

(1) The fields satisfy the WEC, $R_{\alpha\beta}l^{\alpha}l^{\beta}\geq0%
\Rightarrow\rho+p\geq0$;

(2) there are no closed timelike loops and

(3) there exists at least one trapped surface $S_{T}$,

Geodetic incompleteness then leads to the existence of a singularity.

In our treatment we show that for the spacetime in consideration the trapped
region is also bounded from inside so that for $0<r<$ $r_{0}$ one finds $%
\theta ^{+}\theta ^{-}<0$. This makes such a\ spacetime geodetically
complete (no cusping of the in-going and out-going geodesics) which violates
some of the above conditions for the existence of singularity. This is a key
result of the paper.

\section{The spacetime geometry}

\subsection{The Metric}

In this work we investigate whether the spacetime geometry consequent to
gravitational collapse can support trapped surfaces without creating
singularities. We assume that the final collapse has resulted in the
formation of an event horizon that envelopes the matter fields. Then the
real and apparent horizons coincide to locate the \textit{outer} marginally
trapped surface. Further, for simplicity, we assume the end-product to admit
static, spherically symmetric solutions. In Schwarzschild coordinates the
line element for the spacetime can then take the form 
\begin{equation}
ds^{2}=-A\left( r\right) dt^{2}+A\left( r\right) ^{-1}dr^{2}+r^{2}d\Omega
^{2},  \label{metricgen}
\end{equation}%
where $A\left( r\right) =1-\frac{2m\left( r\right) }{r}$ and here the mass $%
m\left( r\right) =4\pi \int_{0}^{r}\rho \left( r^{\prime }\right) {r}%
^{\prime \,2}dr^{\prime }$ is a function of the radial coordinate and is
distributed in some region $r<2m\left( r\right) $. Such a line element is
general enough to contain our spacetimes of interest. There are two boundary
requirements for the spacetime of our interest:

(i) The first is that for large radial distance $r_{M}<r\leq \infty $ the
spacetime be asymptotically Schwarzschild 
\begin{equation}
ds^{2}=-\left( 1-\frac{2M}{r}\right) dt^{2}+\left( 1-\frac{2M}{r}\right)
^{-1}dr^{2}+r^{2}d\Omega ^{2}  \label{metricS}
\end{equation}%
Here $M=4\pi \int_{0}^{\infty }\rho \left( r^{\prime }\right) {r}^{\prime
\,2}dr^{\prime }$ is the total mass and $r_{M}$ gives the surface of the
matter fields, with $r_{M}<2M$.

(ii) The second requirement is that the spacetime be asymptotically de
Sitter 
\begin{equation}
ds^{2}=-\left( 1-\frac{r^{2}}{r_{0}}\right) dt^{2}+\left( 1-\frac{r^{2}}{%
r_{0}}\right) ^{-1}dr^{2}+r^{2}d\Omega ^{2}  \label{metricDeS}
\end{equation}%
for $0\leq r\leq \left( r_{0}<r_{M}\right) $. Here $r_{0}\ $signals the
onset of de Sitter behavior and is given by $r_{0}=\sqrt{3/\Lambda }=\sqrt{%
\frac{3}{\kappa \rho _{0}}}$, where $\rho _{0}=\rho \mid _{r\longrightarrow
0}$ is the upper-bound on the density of the fields and $\kappa =8\pi G$.

The two boundary requirements in Eqs. (\ref{metricS}) and (\ref{metricDeS})
imply that there is an interior region given by $r_{0}\leq r\leq r_{M}$ for
which $m=m\left( r\right) $. The entire spacetime must therefore satisfy
regularity conditions at the two interfaces $r=r_{M}$ and $r=r_{0}$. These
conditions guarantee (i) continuity of the mass function and (ii) continuity
of the pressure across the interfacing hyperfaces. Thus at each interface we
must have for $A(r)$ that 
\begin{equation}
\left[ A^{+}-A^{-}\right] _{\mid r=r_{i}}=0,
\end{equation}%
and 
\begin{equation}
\left[ \partial _{r}A^{+}-\partial _{r}A^{-}\right] _{\mid r=r_{i}}=0,
\end{equation}%
where $r_{i}=\left\{ r_{m},r_{0}\right\} $ and $^{+}$, $^{-}$ refer to the
exterior and interior values respectively.

It will be convenient to transform (Eq. \ref{metricgen}) into coordinates
suitable for discussing null geodesics and which cover all the regions of
interest\footnote{%
The approach is generalization of the regularization of the Schwarzschild
coordinates for a coordinate dependent mass function.}. Note that in this a
spacetime, null fields propagate along geodesics given by 
\begin{equation}
g_{\alpha \beta }k^{\alpha }k^{\beta }=0=-(1-\frac{2m(r)}{r})\left( \frac{dt%
}{d\lambda }\right) ^{2}+(1-\frac{2m(r)}{r})^{-1}\left( \frac{dr}{d\lambda }%
\right) ^{2}  \notag
\end{equation}%
%
%
%
%
%
%
%
%
%
%$,
where $\lambda $ is an affine parameter. The radial geodesics of a massless
particle are therefore given by %$%
\begin{equation}
dt=\pm \left( 1-\frac{2m\left( r\right) }{r}\right) ^{-1}dr=dr^{\ast },~~%
\mathrm{where}~~r^{\ast }=\pm \int \frac{dr}{1-\frac{2m\left( r\right) }{r}}
\end{equation}%
%
%
%
%
%
%
%
%
%
%$
is the a Regge-Wheeler-like tortoise coordinate in this spacetime and $dt\pm
dr^{\ast }=0$. One can set up an Eddington-Finkelstein-like double null
coordinate system $\left\{ u,v\right\} $ given by $u=t-r^{\ast },\ -\infty
<u<\infty $ and $v=t+r^{\ast },\ -\infty <v<\infty $. Note that towards the
future, the radial coordinate increases along constant $u$\ and decreases
along $v$ so that $u$ and $v$ are out-going and in-going null coordinates,
respectively. The coordinates cover the exterior $r>2m\left( r\right) $
region and the metric (Eq. 4) now takes the form 
\begin{equation}
ds^{2}=-A\left( r\right) dudv+r^{2}d\Omega ^{2}.
\end{equation}%
In order to regularize the metric in Eq. (\ref{metricDeS} at the points $%
A\left( r\right) =0$ it is necessary to choose new coordinates $U\left(
u\right) $ and $V\left( v\right) $ ( %which keep the transformations
$v\mapsto U$ and $v\rightarrow V$ which keep the metric invariant), so that 
\begin{equation}
ds^{2}=-\Phi ^{2}\left( u,v\right) dUdV+r^{2}\left( U,V\right) d\Omega ^{2},
\label{metricnull}
\end{equation}%
where the function $\Phi ^{2}\left( u,v\right) $ can be determined and will,
in general, depend on $r$ both explicitly and implicitly through the mass
function $m\left( r\right) $. In comparing Eq. (\ref{metricnull}) with Eq. (%
\ref{metricgen}) one notes the transformation from $r,t$ to $U,V$
coordinates implies that 
\begin{align}
\left( \partial _{t}U\right) \left( \partial _{t}V\right) \Phi ^{2}&
=A\left( r\right) ,  \notag  \label{defA} \\
\left( \partial _{r}U\right) \left( \partial _{r}V\right) \Phi ^{2}&
=-A\left( r\right) ^{-1} \\
\left[ \left( \partial _{r}U\right) \left( \partial _{t}V\right) +\left(
\partial _{r}V\right) \left( \partial _{t}U\right) \right] \Phi ^{2}& =0 
\notag
\end{align}

The three equations in Eq. (\ref{defA}) are identically satisfied by the
choice $U\left( u\right) =-Be^{-\frac{1}{2}\gamma u}$ and $V\left( v\right)
=Be^{\frac{1}{2}\gamma v}$, where $B$ is some scale factor which we can set
to unity, and the $\gamma $s (for the different regions in the spacetime)
are chosen so that $\Phi ^{2}$ is regular at all zeros of $A\left( r\right) $%
. With this choice one finds with use of Eq. (\ref{defA}) that 
\begin{equation}
\Phi ^{2}=\frac{A\left( r\right) }{4B^{2}\gamma ^{2}}e^{-\frac{\gamma }{2}%
\left( v-u\right) }=\frac{A\left( r\right) }{4B^{2}\gamma ^{2}}e^{-\gamma
r^{\ast }},  \label{nextEq}
\end{equation}%
where the parameter $\gamma $ may in general depend on the mass $m\left(
r\right) $ enclosed at the coordinate $r$.

\subsection{ Focusing}

Consider now the spacelike subspace $\Sigma $ of the above spacetime. One
can foliate $\Sigma $ with a family of 2-surfaces $S$ of constant $U$ and $V$%
, which are null hyperfaces ($U$ and $V$ are null coordinates of Eq. \ref%
{metricnull}). Note that $r$ is constant on each such surface $S$. One can
then define radial $\left( d\theta =d\varphi =0\right) $ vectors $l^{\pm }$
tangent to the null geodesics orthogonal to $S$ and which are future
pointing. The ingoing and outgoing vectors are respectively given by $%
^{+}l=\Phi ^{-2}\partial _{V}$, and $^{-}l=\Phi ^{-2}\partial _{U}$ and
satisfy the orthogonality condition $\ g_{\alpha \beta }l^{\alpha }l^{\beta
}=0$. We now consider the divergence $\theta $ of the null geodesics of the
spacetime given by Eq. (\ref{metricnull}). Their general evolution is
depicted in Eq. 2. Let $^{-}l^{\alpha }=(0,a,0,0)$ be a tangent vector to an
outgoing geodesic. By definition, the null geodesics $x^{\alpha }\left(
\lambda \right) $\ with affine parameter $\lambda $, for which the tangent
vectors $\frac{dx^{\alpha }}{d\lambda }$\ are normal to a null hyperface $%
S_{N}$, are the generators of $S_{N}\ $(see e.g. [29]. Then $l^{\alpha
};_{\beta }l^{\beta }=0\Rightarrow a=\Phi ^{-2}$ and the expansion $\theta =$
$l_{~;\alpha }^{\alpha }$ is, for in-going null geodesics given by 
\begin{equation}
\theta ^{+}=\frac{2}{r\Phi \left( r\right) ^{2}}\frac{\partial r}{\partial V}%
.,  \label{thetapl}
\end{equation}%
and for out-going null geodesics, by 
\begin{equation}
\theta ^{-}=\frac{2}{r\Phi \left( r\right) ^{2}}\frac{\partial r}{\partial U}%
.  \label{thetamin}
\end{equation}

\section{A Non-Singular Spacetime}

The Dymnikova solution [30] depicts a plausible scenario for a spacetime
that could result from gravitational collapse. In this solution the fields $%
T_{\mu \nu }$ satisfy the conditions 
\begin{equation}
T_{t}^{t}=T_{r}^{r};\ T_{\theta }^{\theta }=T_{\varphi }^{\varphi }
\end{equation}%
The radial pressure $p_{r}$ satisfies a cosmological constant-like equation
of state $p_{r}=-\rho $, while the tangential pressure satisfies $p_{\theta
}=p_{\varphi }=-\rho -\frac{1}{2}r\frac{\partial \rho }{\partial r}$. The
analysis [30] gives an exact solution to the Einstein equations which takes
the form of Eqs. \ref{metricgen} - \ref{metricDeS} with a mass function of
the form 
\begin{equation}
m\left( r\right) =4\pi \int_{0}^{x}\rho \left( x\right) x^{2}dx=M\left[
1-e^{-\left( \frac{r^{3}}{r_{0}^{2}r_{g}}\right) }\right] ,  \label{massfun}
\end{equation}%
where $m\left( r\right) $ is the mass enclosed within a radius $r$. The
radial density profile $\rho \left( r\right) $ is given by $\rho \left(
r\right) =\rho _{0}e^{-\left( \frac{r^{3}}{r_{0}^{2}r_{g}}\right) };$ with \ 
$r_{0}=\frac{3}{\Lambda }$ and $r_{g}=2MG$.

\subsection{Regularized metric}

Notice that the Dymnikova solution contains no matter fields (in the usual
sense) since the only field existing has a vacuum-like equation of state in
the radial direction. This allows one to seek a Kruskal-like (Eq. \ref%
{metricnull}) extension of this spacetime. We are only interested in
regularizing the coordinate singularities in Eqs. \ref{metricgen} - \ref%
{metricDeS} in order to follow the expansion $\theta $. In the region $0\leq
r\leq \infty $ this spacetime can be divided into three sub-regions.

\underline{Region 1}: $r_{M}<r\leq \infty $. This is the Schwarzschild
vacuum region. Here $r_{M}<2M$ describes the surface of matter fields and is
located deep inside the Schwarzschild horizon. It follows that the the
problem of satisfying the junction conditions (for matter fields) across the
horizon (see e.g. [1]) does not arise. The regularization of the coordinates
at $r=2M$ can then be effected. Thus Region 1 includes two sectors of the
Schwarzschild vacuum: (1a) the exterior $2M<r\leq \infty $ and (1b) the
interior $r_{M}<r\leq 2M$. In region 1a) we choose [31] $U<0,V>0$, $\gamma
_{1}=\frac{1}{4M}$ and $A=\frac{1}{2}$. In general for Region 1 we have that
the metric is same as in Eq.11 with 
\begin{equation}
\Phi ^{2}=\frac{32M^{3}}{r}e^{-\frac{r}{2M}},\ r_{M}<r\leq \infty .
\end{equation}%
In the present work we shall squash region 1b so that $r_{M}<r\leq 2M$
consists of only the inner 2-surface $r_{M}^{-}$, and region 2 containing
the anisotropic fluid of Eq. starts here\footnote{%
As is shown from the form of the mass function Eq. 17, the boundary at $%
r_{M}^{-}$ is a soft boundary which easily satisfies the conditions in Eqs.
7 and 8.}.

\underline{Region 2}: $r_{0}<r<r_{M}$. This is the intermediate interior
region containing containing a field with a radial dependence $m\left(
r\right) $ as in Eq. 16 and a vacuum-like equation of state $p_{r}\left(
r\right) =-\rho \left( r\right) $. Here (as in region 1b) we have $U>0,$ $%
V>0 $ and the metric is still given by Eq. 10. Since however $r^{\ast }$
does not admit a simple closed form integral and one cannot determine $%
\gamma $ explicitly. As it turns out this is not a serious problem since the
region $r_{0}<r<r_{M}$ has no coordinate or physical singularities. Thus in
following $\theta $ in this region, we only require to match region to
regions 1 and 3 using the matching conditions already mentioned above. 
\begin{equation}
\Phi ^{2}=\frac{A\left( r\right) }{4B^{2}\left( \gamma _{2}\right) ^{2}}%
e^{-\gamma _{2}r^{\ast }\left[ =-\frac{\gamma }{2}\left( v-u\right) \right]
},\ r_{0}<r\leq r_{M}  \label{phi2}
\end{equation}%
.

\underline{Region 3:} $0\leq r\leq r_{0}$ contains an (almost) constant
density $\rho _{0}$ vacuum-like fluid of density $\Lambda $ with a length
scale given by $r_{0}^{2}=\frac{3}{\Lambda }$. The solution with the mass
function of Eq. (\ref{massfun} has an inner horizon $r_{-}$ at $r_{0}$
(actually at $1.00125\,r_{0}$) provided the total collapsed mass $M$ is
large enough, i.e. $M>m_{crit}\simeq 0.3m_{Pl}\sqrt{\frac{\rho _{Pl}}{\rho
_{0}}}$. Thus to maintain regularity of the metric across $r=r_{0}$ we
introduce $U>0$ and $V<0$. Here the metric becomes asymptotically de Sitter,
taking the form of Eq. (\ref{metricDeS}) with $A\left( r\right) \sim \left(
1-\frac{r^{2}}{r_{0}}\right) $, $0<r<r_{0}$. This gives $r^{\ast }=\int
\left( 1-\frac{r^{2}}{r_{0}}\right) ^{-1}dr=\frac{r_{0}}{2}\ln \frac{r_{0}+r%
}{r_{0}-r}$. So that now 
\begin{equation}
\Phi ^{2}=\frac{A\left( r\right) }{4B^{2}\left( \gamma _{3}\right) ^{2}}%
e^{-\gamma _{3}r^{\ast }}=\left( \frac{1}{r_{0}}\right) ^{2}\left(
r_{0}+r\right) ^{1-\gamma _{3}r_{0}}\left( r_{0}-r\right) ^{1+\gamma
_{3}r_{0}},\ 0\leq r\leq r_{0}.
\end{equation}%
By inspection of Eq. (\ref{phi2}), it follows that to make $\Phi ^{2}$
regular in this region one requires that $1+\gamma _{3}r_{0}=0$ giving $%
\gamma _{3}=-\frac{1}{r_{0}}$.

\begin{figure}[tbh]
\centerline{\epsfig{file=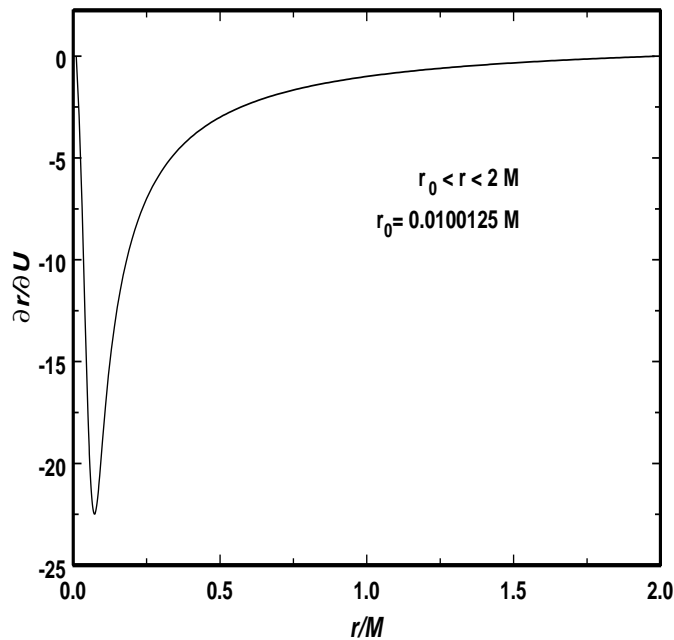,height=10.0cm,width=10cm}}
\caption{{\protect\small The $\partial r / \partial U$ derivative (to which
the expansion $\protect\theta$ is proportional) for the region between the
outer and inner horizons i.e. $r_{_{0}} < r < 2M$ with $r_{_{0}}= 0.00100125
\, M $ the position of the inner horizon. Instead of reaching $- \infty$ as $%
r$ approaches zero, the expansion $\protect\theta$ begins increasing for $r
> (r_{_{0}}^{2} r_{g})^{1/3}$ approaching zero at the horizon $r = r_{_{0}}$.%
}}
\end{figure}

\begin{figure}[tbh]
\centerline{\epsfig{file=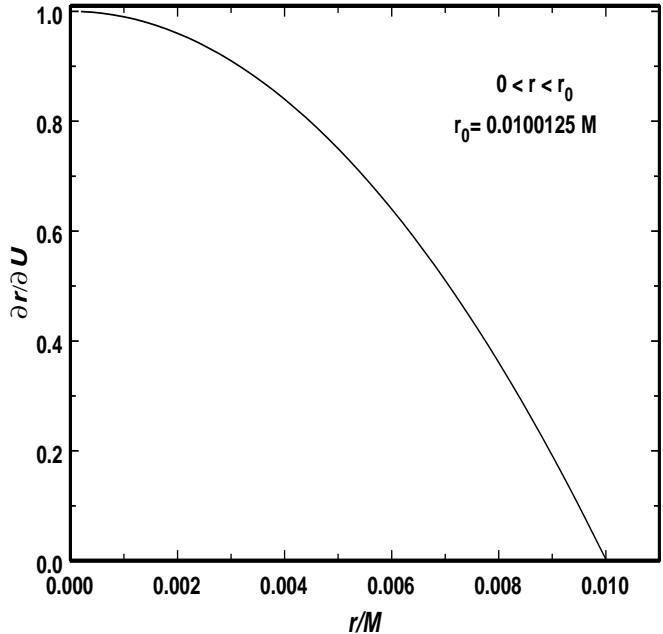,height=10.0cm,width=10cm}}
\caption{{\protect\small Same as figure 1 but for the region interior to the
inner horizon $0 < r < r_{_{0}}$. The expansion vanishes on the horizon
while it reaches its maximum value at $r = 0$. When plotted in a single
figure this curves joins smoothly that of figure 1 at the inner horizon.}}
\end{figure}

\subsection{Null geodesic convergence in Dymnikova spacetime}

Using the results in Eqs. \ref{massfun}-\ref{phi2} we can now compute the
expansions of the in-going $\theta ^{+}=\frac{2}{r\Phi \left( r\right) ^{2}}%
\frac{\partial r}{\partial V}$ and out-going $\theta ^{-}=\frac{2}{r\Phi
\left( r\right) ^{2}}\frac{\partial r}{\partial U}$ null geodesics in the
above spacetime for the entire region $0\leq r<\infty $. Fig. 1 is a plot of 
$\partial r/\partial U$ (to which $\theta $ is proportional) for the
outgoing mode in this spacetime in the region $r_{0}<r<2M$.

\bigskip In the region $r>2M$, the geodesic congruences expand in a similar
way as they do in the Schwarzschild spacetime. Thus here $\theta ^{+}>0$ and 
$\theta ^{-}<0$, so that $\theta ^{+}\theta ^{-}<0$. However as one
traverses the region $r<2M$, the dependence of $\theta $ on $r$ evolves very
differently. In the Schwarzschild spacetime the out-going null congruences
are given by $\theta ^{+}<0$ and evolve such that $\theta ^{+}\rightarrow
-\infty $ as $r\rightarrow 0$. But in this region we still have for the
in-going congruences that $\theta ^{-}<0$. Thus it is expected that
eventually the two congruences would intersect or cusp leading to an
incomplete spacetime. This signifies the existence of a physical singularity
[24] [26]. One can compare this situation with that for the spacetime under
consideration. One notices that in the region $r<2M$ of the Dymnikova
spacetime (see Fig. 1) the expansion of the outgoing mode $\theta ^{+}$
starts off decreasing as in the Schwarzschild case. However, near $r\simeq
(r_{0}^{2}\,r_{g})^{1/3}$ it reaches a minimum and and thereafter, for
smaller radii, $\theta ^{+}$ begins increasing as the metric begins
deviating significantly from that of Schwarzschild. Eventually, at $r=r_{0}$
(in reality at $r=1.00125\,r_{0}$), $\theta ^{+}$ vanishes, to signify the
presence an inner marginally trapped surface (and hence horizon) at this
point. In figure 2 we show the function $\partial r/\partial U$ for points
in the region $r<r_{0}$. One notices that $\theta $ is positive for outgoing
modes in this region, as indeed is the case in the $r>2M$ region, but
vanishes at $r=r_{0}$, which is a marginally trapped surface i.e. a horizon.

In analyzing spacetimes based on the behavior of its null congruences $%
\theta ^{+}\ $we have relied mostly on the behavior of the out-going
geodesics. A more recent approach \cite{[32]} (which we shall also apply) is
to consider the evolution of $\theta ^{+}\theta ^{-}$ with the radial
coordinate $r$. This is a more general (and in our case more straight
forward) approach. When $\theta ^{-}$ remains converging (as in the
Schwarzschild case) then it is easy to verify that the sign of $\theta
^{+}\theta ^{-}$ follows the sign of $\theta ^{+}$. Thus $\theta ^{+}\theta
^{-}<0$ signifies a regular spacetime, $\theta ^{+}\theta ^{-}=0$ signifies
a marginally trapped surface and $\theta ^{+}\theta ^{-}>0$ signifies a
trapped region. In a maximally extended spacetime, Eqs. 14 and 15 imply that
the product of the expansions is given by $\theta ^{+}\theta ^{-}=\frac{4}{%
r^{2}\Phi ^{2}\left( r\right) }\frac{\partial r}{\partial U}\frac{\partial r%
}{\partial V}$. On using our definitions of $U\left( u\right) $ and $V\left(
v\right) $ to Eqs. 12 we find \ 
\begin{equation}
\theta ^{+}\theta ^{-}=-\frac{4}{r^{2}}\left[ 1-\frac{2m(r)}{r}\right]
\end{equation}%
with $m(r)\ $given by Eq. (\ref{massfun}).\ The expression $-(r^{2}/4)\theta
^{+}\theta ^{-}$ is plotted against $r$ in Figure 3. The plot shows (as
expected) that $\theta ^{+}\theta ^{-}$ first changes sign at the outer
horizon where it turns from negative to positive as it enters the trapped
region $r<2M$. Inside the trapped region $\theta ^{+}\theta ^{-}$ approaches
some maximum and then decreases to zero at some value of $r$, signifying an
exit out of the trapped region. Thereafter, $\theta ^{+}\theta ^{-}$ grows
without bound in the negative direction. It is notable that $\theta
^{+}\theta ^{-}$ remains bounded in the trapped region $r_{0}<r<2M$ and does
not tend to $+\infty $. Further, it is notable that $\theta ^{+}\theta
^{-}<0 $ in the region $0\leq r<r_{0}$, just like in the region $%
2M<r<+\infty $. It follows that the entire spacetime under consideration is
geodetically complete. This violates a condition (in the singularity
theorems) for the formation of a matter-generated future-directed
singularity. We therefore have an example of a trapped region with no
singularity. This finding was the purpose of our inquiry.

\begin{figure}[tbh]
\centerline{\epsfig{file=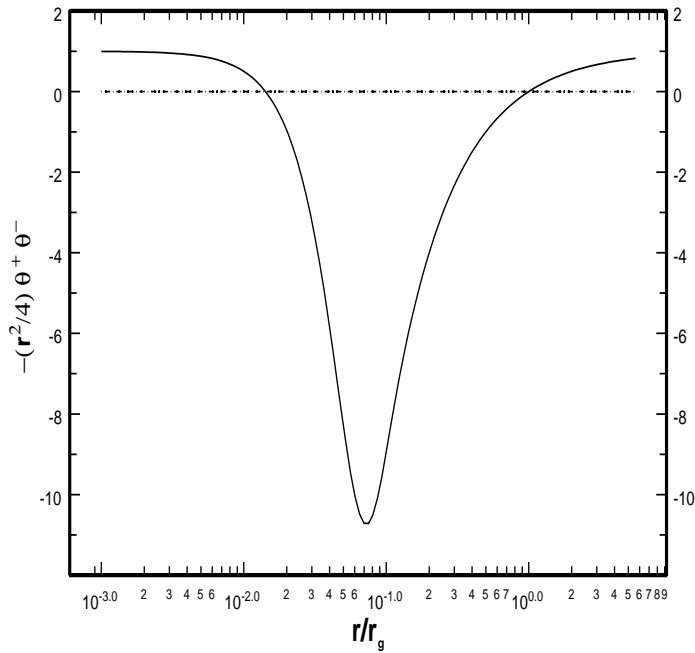,height=10.0cm,width=10cm}}
\caption{{\protect\small The expression $-(r^{2}/4)\protect\theta ^{+}%
\protect\theta ^{-}$ as a function of the normalized radial coordinate $%
r/r_{g}$. It is apparent that the quantity $\protect\theta ^{+}\protect%
\theta ^{-}$ changes sign at the inner horizon becoming negative for $%
r<r_{0} $.}}
\end{figure}

\section{Conclusion}

In this work we have investigated the question of whether in gravitational
collapse spacetime can sustain trapped surfaces without a singularity. As a
concrete example we have considered the Dymnikova spacetime, which is \ a
static spherically symmetric solution of the Einstein Field Equations with a
de Sitter-like fluid at the core. The investigations performed by following
the expansion $\theta $ of the null geodesic congruences of the spacetime as
a function of the radial coordinate. It is found that for large values of $r$%
, i.e. $\infty <r\leq 2M$ the behavior of out-going null $\theta ^{+}$ and
that of the product $\theta ^{+}\theta ^{-}$ are similar to those of the
Schwarzschild space, in that $\theta ^{+}\geq 0$ and $\theta ^{+}\theta
^{-}\leq 0$ and vanishing at the horizon boundary, $r\leq 2M$. As one
traverses the region $r<M$ of the Dymnikova spacetime one find that,
initially $\theta ^{+}<0$ and is growing more negative and $\theta
^{+}\theta ^{-}<0$. In the Schwarzschild spacetime this behavior leads to
geodetic incompleteness and the occurrence of a singularity. In the
Dymnikova spacetime, however, both $\theta ^{+}$ and $\theta ^{+}\theta ^{-}$
have turning points (minimum and maximum, respectively) in this region and
eventually both vanish around $r=r_{0}$. For $0\leq r<r_{0}$ one finds that $%
\theta ^{+}>0$ and $\theta ^{+}\theta ^{-}<0$.

\ \ \ \ \ \ The main results are that in this spacetime, there is a trapped
region with an outer and inner boundary. In this trapped region, null
congruences do not converge without bound and both $\theta ^{+}$ and $\theta
^{+}\theta ^{-}$ have a minimum. Further, there is a region $0\leq r<r_{0}$
including the black hole center $r=0$, in which the spacetime is regular,
with converging in-going congruences $\theta ^{-}<0$ and diverging out-going
congruences, $\theta ^{+}>0$ so that $\theta ^{+}\theta ^{-}<0$. These
results suggest that the trapped region $\Sigma _{T}$: (a) does not enclose
a singularity and, (b) may not necessarily include the origin.

In \cite{[33]} we have constructed a solution to the Einstein Field
Equations for a static spherically symmetric space-time with an equation of
state for a gravitating fluid that transits smoothly between that of a
matter-like fluid ($P=w\rho ,\ $ $w>0$) and evolves to vacuum-like state, ($%
w=-1$). While a maximal extension of this spacetime is still to be made one
expects, based on the general arguments above and pending verification, that
its trapped surfaces will too be non-singular. Generally, these arguments
suggest that if gravitational collapse were to proceed with some phase
transition that leads to the creation of some de Sitter-like fluid $\left(
p=-\rho \right) $ at the collapse core deep inside the Schwarzschild surface
then it is plausible that a trapped region with no singularity would result
inside the black hole.

\end{document}